\documentclass[12pt,epsf]{article}
\usepackage{amsmath}
\usepackage{amssymb}
\usepackage[nosort]{cite}
\newcommand{\CC}{{\cal C}}

\newcommand{\OO}{{\cal O}}

\newcommand{\pint}{\makebox[0pt][l]{\hspace{3.4pt}$-$}\int}

\newcommand{\be}{\begin{equation}}
\newcommand{\ee}{\end{equation}}
\newcommand{\ben}{\begin{eqnarray}\displaystyle}
\newcommand{\een}{\end{eqnarray}}

\newcommand{\bw}{\overline{w}}
\newcommand{\s}{\sigma}
\newcommand{\ts}{\tilde{\sigma}}
\newcommand{\la}{\lambda}

\newcommand{\Tr}{\hbox{Tr}}

\newcommand{\sectiono}[1]{\section{#1}\setcounter{equation}{0}}

\begin{document}

{}~ \hfill\vbox{\hbox{hep-th/0310188}
\hbox{UUITP-19-03}\break
\hbox{CTP-MIT-3425}\break
\hbox{ITEP-TH-53/03}}\break

\vskip 1.cm

\centerline{\large \bf Yang-Mills Duals for Semiclassical Strings
on $\bf AdS_5\times S_5$}
\vspace*{5.0ex}

\centerline{\large \rm J.~Engquist${}^a$, J.~A.~Minahan${}^{a,b}$ and 
K.~Zarembo${}^a$\footnote{Also at ITEP, Bol. Cheremushkinskaya 25, Moscow, Russia}}

\vspace*{2.5ex}
\centerline{\large \it${}^a$ Department of Theoretical Physics}
\centerline{\large \it Box 803, SE-751 08 Uppsala, Sweden}
\vspace*{2.5ex}

\centerline{\large \it${}^b$ Center for Theoretical Physics}
\centerline{\large \it Massachusetts Institute of Technology}
\centerline{\large \it Cambridge, MA 02139, USA}
\vspace*{2.5ex}

{\tt johan.engquist,joseph.minahan,konstantin.zarembo@teorfys.uu.se}\\[20mm]

\centerline {\bf\large Abstract}

\bigskip\bigskip
We consider a semiclassical multiwrapped circular string pulsating
on $S_5$, whose center of mass has angular momentum $J$ 
on an $S_3$
subspace.  Using the AdS/CFT correspondence we argue that
the one-loop anomalous dimension of the dual operator is
a simple rational function of $J/L$, where $J$ is the $R$-charge
and $L$ is the bare dimension of the operator. 
  We then reproduce this result directly from a super Yang-Mills
computation, where we make use of the integrability of the one-loop
system to set up an  integral equation that  we 
solve.  We then verify the results of 
Frolov and Tseytlin for circular
rotating strings with $R$-charge assignment $(J',J',J)$.
In this case we solve for an integral equation found in
the $O(-1)$ matrix model when $J'< J$ and the $O(+1)$ matrix model if
$J'> J$.  The latter region starts at $J'=L/2$ and continues down,
but an apparent critical point is reached at  $J'=4J$.  We argue
that the critical point is just an artifact of the Bethe ansatz and
that the conserved charges of the underlying integrable model are
analytic for all $J'$ and that the results from the $O(-1)$ model
continue onto the results of the $O(+1)$ model.

\bigskip

\vfill \eject
\baselineskip=17pt

\sectiono{Introduction}

While the AdS/CFT conjecture \cite{Maldacena:1998re,Gubser:1998bc,Witten:1998qj} is generally accepted as fact,
proving it is a highly nontrivial problem.  
However, there are
many nontrivial tests that can be applied to the conjecture,
and these tests may provide insight toward a formal proof.
Furthermore, working out the consequences of the AdS/CFT duality in concrete examples
uncovers a beautiful interplay between quantum fields and strings.

The most celebrated ``stringy'' test of the conjecture is the 
comparison of
the string spectrum in the plane-wave limit \cite{Metsaev:2001bj,Metsaev:2002re}
to the anomalous dimension
of single trace operators
that differ from BPS protected operators by a finite amount 
\cite{Berenstein:2002jq}.
More recently, a program
has begun comparing the spectrum of semiclassical string configurations
\cite{Gubser:2002tv,Frolov:2002av,Russo:2002sr,Minahan:2002rc,Tseytlin:2002ny,
Frolov:2003qc,Frolov:2003tu,Frolov:2003xy,Arutyunov:2003uj}
with  anomalous dimensions 
for a wider class of operators \cite{Minahan:2002ve,Belitsky:2003ys,Beisert:2003xu,Beisert:2003jj,
Beisert:2003yb,Beisert:2003ea,Gorsky:2003nq}.  
These operators  also have large charges,
but are not necesserily close to BPS operators.

One such comparison was made between a folded string in
$S_5$, rotating in one plane and revolving in another 
\cite{Frolov:2003xy,Arutyunov:2003uj},
 and a single trace operator composed of  two types of scalar fields
\cite{Beisert:2003xu}. 
The energy of the string corresponds to the scaling dimension
of an operator on the gauge theory side, and this
was found by mapping the problem to the Heisenberg spin chain, where
the anomalous dimension can be computed as an
eigenvalue of the spin Hamiltonian \cite{Minahan:2002ve}.
The integrability of the Heisenberg model is a powerful tool, allowing one to
reduce the problem to solving a series of Bethe equations \cite{Bethe,faddeev}.  
The semiclassical limit of the string corresponds to the
thermodynamic limit in the spin system, which
allows one to
convert the Bethe equations into a relatively simple integral equation.
This has a striking resemblance to saddle-point equations 
in certain large-$N$ matrix models and  is solved by similar techniques.
The result is a parametric 
relationship between the
anomalous dimension and the R-charges of the Super Yang-Mills (SYM) operator
\cite{Beisert:2003xu}.
The same relationship  between the energy and the angular momenta
arises in the string computation \cite{Frolov:2003xy,Arutyunov:2003uj}.
This result was generalized to folded strings that rotate  not only in $S_5$, but 
also in the $AdS_5$ \cite{Beisert:2003ea}, 
again showing a beautiful agreement between string theory
and  SYM.

The string states on $S_5$ and the scalar operators in the SYM theory can be
characterized by three R-charges  which define the highest weight $(J_1,J_2,J_3)$
of an $SO(6)$ representation with Dynkin indices $[J_2+J_3,J_1-J_2,J_2-J_3]$.
On the string side, 
the R-charges correspond to the
angular momenta on $S_5$. The simplest operator in the SYM theory
with the R-charge assignment $(J_1,J_2,J_3)$ is $\Tr\,( X^{J_1}Y^{J_2}Z^{J_3})$, where 
$X$, $Y$ and $Z$ are the standard complex scalars of the ${\cal N}=4$
supermultiplet,
\be
X=\frac{1}{\sqrt{2}}(\phi_1+i\phi_2)\qquad Y
=\frac{1}{\sqrt{2}}(\phi_3+i\phi_4)\qquad Z=\frac{1}{\sqrt{2}}(\phi_5+i\phi_6)
\ .
\ee
The R-charge does not depend on the ordering in the trace but
operators with different ordering mix under renormalization.
 This mixing  makes computation of
one-loop anomalous dimensions a nontrivial problem. The efficient way to resolve the
operator mixing is to map the problem to the Hamiltonian of an
integrable $SO(6)$ spin chain \cite{Minahan:2002ve}, 
which is then solved using the Bethe ansatz.
The bare dimension of a holomorphic operator, which does not contain
$\overline{X}$, $\overline{Y}$ or $\overline{Z}$, is a sum of the R-charges. This property resembles 
BPS saturation, and indeed some of the 
string solitons of 
\cite{Frolov:2003qc,Frolov:2003tu,Frolov:2003xy,Arutyunov:2003uj} are supersymmetric in the tensionless limit of string
theory \cite{Mateos:2003de}, the limit that is supposed to be dual to free SYM.
 We strongly believe this asymptotic supersymmetry is important but not
really necessary for establishing the precise agreement between semiclassical
string states and  SYM operators. To support this point of view, we will 
give examples of  string states that cannot satisfy a BPS bound by dimensional counting.
We will construct AdS duals of these states in a way similar to the
construction of   duals for the
asymptotically supersymmetric states. The corresponding 
SYM operators will  contain both holomorphic and anti-holomorphic fields,
therefore, their bare dimension will {\it exceed} their
R-charge by an arbitrarily large amount. 

Plenty of string solutions with all possible charge assignments are known
\cite{Arutyunov:2003uj}.
Some of these solutions predict a simple analytic relation between
the anomalous dimension and the $R$-charges or the bare dimension.
We will concentrate on the duals of these string states.

For example, the folded string state of \cite{Frolov:2003xy,Arutyunov:2003uj}
has an SYM dual with $R$-charge assignment $(J',0,J)$.  The string
prediction was shown to match with the one-loop SYM result, but the
relation between the anomalous dimension and the ratio of $J'/J$ involved
elliptic functions \cite{Frolov:2003xy,Beisert:2003xu,Beisert:2003ea}.   
When $J'=J$,  the folded string has the same $R$-charges as a circular
string that wraps around two planes and is stationary in a third
\cite{Frolov:2003qc,Frolov:2003tu}.
In \cite{Beisert:2003xu} 
the SYM dual for this string was found, where not
only the anomalous dimension was shown to be the same, but the fluctutation
spectrum was also shown to match.  But it is also simple to compute the
energy of  a circular string when there {\it is} a center of mass
motion in the third plane, and the $R$-charges are of the
form $(J',J',J)$.  
Expanding about small 't Hooft coupling $\lambda=g^2_{YM}N$, one finds that
the corresponding anomalous dimension has the particularly simple form
\be\label{gamma_J'J'J}
\gamma\ =\ \frac{m^2\lambda J'}{L^2}
\ee
where $L=2J'+J$ and $m$ counts the string winding.
Given this extraordinarily simple equation, it seems likely that
such a result can be reproduced by a one-loop SYM computation.  One
of the goals of this paper is do precisely this.

We will also identify an even simpler duality. 
We consider a string that is pulsating between two extremal points on
$S_5$,
where the string also has center of mass motion along an $S_3$
subspace, giving it angular momentum $J$.   This is a generalization
of a case previously considered in \cite{Minahan:2002rc}.  Using simple first
order perturbation theory, we will show that the prediction for the
one-loop anomalous dimension is
\be\label{gamma_00J}
\gamma\ =\ \frac{m^2\lambda (L^2-J^2)}{4L^3},
\ee
We then reproduce this equation for a single trace operator
with $R$-charge assignment $(0,0,J)$ and 
bare dimension $L$.  It is not necessary that  $J$  be close to $L$.

{}From the SYM point of view, we will see that  equations
\eqref{gamma_J'J'J} and \eqref{gamma_00J} are simple because the integral
equations reduce to those found with $O(n)$ matrix models
\cite{Kostov:fy,Gaudin:vx,Kostov:pn,Eynard:1992cn,Eynard:1995nv,Eynard:1995zv,
Chekhov:1996xy}.  For these models, if $n$ is parameterized as
\be
n=2\cos(\pi p/q)
\ee
with $p$ and $q$  positive integers having no common factor, then the resolvant
of the eigenvalue density
is the solution of a polynomial equation of order $q$.   We will see
that the Bethe equations for the dual of the pulsating string reduce
to the integral equation of an $O(0)$  model with a critical
point at $J=0$.   This case is particularly simple since $p/q=1/2$.

For the duals of the circular string, we need to consider two regions.
The first region has $J'<J$ and the Bethe equations reduce to an
$O(-1)$ integral equation.  The second region has $J'>4J$, where after
some work, the Bethe equations are reduced to an $O(+1)$ integral
equation.   The point $J'=4J$ corresponds to a critical point
in the $O(+1)$ model, but we believe that this is only an artifact
of the Bethe ansatz and not a true critical point for the SYM operators.
Indeed, one can analytically continue our results for $J'<J$ up
to $J'=4J$, where the $O(-1)$ model also has a critical point, and
then continue beyond this point to $J=0$.  We find that not only 
is the anomalous dimension analytic across the critical point but
so are all the conserved charges of the underlying integrable model,
and that they match onto the $O(+1)$ results.

The paper is organized as follows.  In section 2 we compute the energy
spectrum for the string pulsating on $S_5$.  In section 3 we consider
the one-loop anomalous dimensions for $R$-charge assignments
$(0,0,J)$  and $(J',J',J)$ with $J'<J$, where the bare dimension 
is $L>J$  in the first case and $L=2J'+J$ in the second.  In sections 4
and 5 we solve the integral equations for these cases and show that
the anomalous dimensions are given by \eqref{gamma_00J} and
\eqref{gamma_J'J'J}.  In section 6
we consider the case with $R$-charge assignment $(J',J',J)$ and
$J'>J$, and show that the anomalous dimension is still given
by \eqref{gamma_J'J'J}.  In section 7 we present our conclusions.

\sectiono{A string pulsating on $S_5$}

In this section we generalize the results of \cite{Minahan:2002rc} 
to include an $R$-charge.
Let us consider a circular pulsating string expanding
and contracting on $S_5$,  which has a center of mass that
is moving on an $S_3$ subspace.  We will assume that the string is
fixed on the spatial coordinates in $AdS_5$, so the relevant metric for
us is
\be\label{S5met}
ds^2\ =\ R^2(-dt^2+\sin^2\theta\ d\psi^2+\ d\theta^2\ +\ \cos^2\theta\ d\Omega_3^2),
\ee
where $d\Omega_3$ is the metric on the $S_3$ supspace and 
$R^2=2\pi\alpha'\sqrt{\lambda}$.
We will assume that the string is stretched along $\psi$ but not
along any of the coordinates in $S_3$.  
If we identify $t$ with $\tau$ and $\psi$ with $m\s$ to 
allow for
multiwrapping, the Nambu-Goto action then 
reduces to
\be\label{NGsph}
S\ =-\ m\sqrt{\la}\int dt\ \sin\theta\ \sqrt{1-\dot\theta^2-\cos^2\theta g_{ij}
\dot\phi^i\dot\phi^j},
\ee
where $g_{ij}$ is the metric on $S_3$ and $\phi^i$ refers to the coordinates
on $S_3$.
Hence, the canonical momentum are
\begin{eqnarray}
\Pi_\theta\ &=&\ \frac{m\sqrt{\la}\ \sin\theta\ \dot\theta}{\sqrt{1-\dot\theta^2-\cos^2\theta g_{ij}\dot\phi^i\dot\phi^j}},\\
\Pi_i\ &=&\ \frac{m\sqrt{\la}\ \sin\theta\ \cos^2\theta g_{ij}\dot\phi^j}{\sqrt{1-\dot\theta^2-\cos^2\theta g_{ij}\dot\phi^i\dot\phi^j}}.
\end{eqnarray}
Solving for the derivatives in terms of the canonical momenta
and substituting into the Hamiltonian, we find 
\be
H\ =\ \sqrt{\Pi_\theta^2+\frac{g^{ij}\Pi_i\Pi_j}{\cos^2\theta}+m^2\la\sin^2\theta}.
\ee
The square of $H$ looks like the Hamiltonian for a particle on $S_5$
with an angular dependent potential.  Since we are interested in large
quantum numbers, the potential may be considered a perturbation.
We thus proceed by considering free wavefunctions on $S_5$ and then
do first order perturbation theory to find the order $\lambda$ correction.
The total $S_5$ angular momentum quantum number will be denoted by
$L$ and the total angular momentum quantum number on $S_3$ is $J$.
Since the potential depends only on $\theta$, we may replace $g^{ij}\Pi_i\Pi_j$
with $J(J+2)$. 

The wave-functions are solutions to the Schr\"odinger equation
\be
L(L+4)\Psi(w)=-\frac{4}{w}\frac{d}{dw}w^2(1-w)\frac{d}{dw}\Psi(w)+
\frac{J(J+2)}{w}\Psi(w),
\ee
where $w=\cos^2\theta$.
In order to simplify the discussion, we will assume that $J$ and $L$ are even
and define $j=J/2$ and $\ell=L/2$.
In this case, the normalized $S_5$ wave functions are given by
\be
\Psi(w)=\frac{\sqrt{2(\ell+1)}}{(\ell-j)!}\frac{1}{w^{j+1}}\left(\frac{d}{dw}\right)^{\ell-j}w^{\ell+j}(1-w)^{\ell-j}.
\ee
Hence, the first order correction to $E^2$ is
\be
\int_0^1wdw\Psi(w)m^2\lambda(1-w)\Psi(w)= m^2\lambda \frac{2(\ell+1)^2-(j+1)^2-j^2}{(2\ell+1)(2\ell+3)}.
\ee
Thus, up to first order in $\lambda$ and assuming $L$ and $J$ large, 
$E^2$ is given by
\be
E^2\ =\ L(L+4)\  +\ m^2\lambda\frac{L^2-J^2}{2L^2}
\ee

Now the bare dimension is $L$, and so the anomalous dimension is
given by
\be\label{anom}
\gamma=\frac{m^2\lambda}{4L}\alpha(2-\alpha),
\ee
where $\alpha=1-J/L$, which we have defined for later convenience.
In the next section we will reproduce this result in a one-loop SYM
computation.

Even though $L$ is nominally a quantum number on $S_5$, 
it is not the $R$-charge.
This is because the wave function chosen is for a rigid string, not a particle.
Any contribution to the total angular momentum for a section of string moving
along $\theta$ is cancelled by the section halfway around the string. Only
the quantum number $J$ on $S_3$ contributes to the $R$-charge since
we are assuming that the string is not stretched along here.

\sectiono{Setting up the gauge theory computations}

In this section, we derive integral equations for the gauge theory
computations that will be solved for in the subsequent two sections.
In order to do these computations, we capitalize on the fact that the 
one-loop anomalous dimension can be mapped to a Hamiltonian of an integrable
spin chain \cite{Minahan:2002ve}.  With this, we can write down a set of Bethe equations that
can be solved in the limit that the number of sites in the chain is large.
We will consider single trace operators $\OO$
 made up of scalar
fields only.  The operators are not required to be holomorphic, but
can contain $\overline{X}$, $\overline{Y}$ and $\overline{Z}$ scalar
fields inside the trace.   We will assume that the operators are highest
weights of  $SO(6)$ representations
and that the $R$-charges
have  the general form $(J',J',J)$.  In terms of $SO(6)$ Dynkin indices, these 
representations are denoted by $[0,J-J',2J']$ if $J'\le J$ and 
$[J'-J,0,J'+J]$ if $J'\ge J$.  If $J'\ne 0$, then we will assume
that the bare dimension $L$ of $\OO$ 
maximizes the BPS-like bound $L=J+2J'$.  If
$J'=0$ we will relax this condition.

In \cite{Minahan:2002ve} 
it was argued that the anomalous dimension of $\OO$ can
be found by solving a series of Bethe equations for a set of Bethe roots.
There are three types of Bethe roots, with each type associated with
a simple root of the $SO(6)$ Lie algebra.  Assuming that there are
$L$ scalar fields in $\OO$, the Bethe equations are given by
\begin{eqnarray}\label{betheSO6}
\left(\frac{u_{1,i}+i/2}{u_{1,i}-i/2}\right)^L&=&
\prod_{j\ne i}^{n_1}\frac{u_{1,i}-u_{1,j}+i}{u_{1,i}-u_{1,j}-i}
\prod_{j}^{n_2}\frac{u_{1,i}-u_{2,j}-i/2}{u_{1,i}-u_{2,j}+i/2} 
\prod_{j}^{n_3}\frac{u_{1,i}-u_{3,j}-i/2}{u_{1,i}-u_{3,j}+i/2}
\nonumber\\
1&=&\prod_{j\ne i}^{n_{2}}\frac{u_{2,i}-u_{2,j}+i}{u_{2,i}-u_{2,j}-i}
\prod_{j}^{n_{1}}\frac{u_{2,i}-u_{1,j}-i/2}{u_{2,i}-u_{1,j}+i/2}
\nonumber\\
1&=&\prod_{j\ne i}^{n_{3}}\frac{u_{3,i}-u_{3,j}+i}{u_{3,i}-u_{3,j}-i}
\prod_{j}^{n_{1}}\frac{u_{3,i}-u_{1,j}-i/2}{u_{3,i}-u_{1,j}+i/2}\,.
\end{eqnarray}
where  $n_1$, $n_2$ and $n_3$ denote the number of Bethe roots associated
with each simple root of $SO(6)$.  For this choice, the Dynkin indices
of this representation
are given by $[n_1-2n_2, L-2n_1+n_2+n_3,n_1-2n_3]$.  
The anomalous dimension is only directly related to the $u_1$ roots 
and is given by
\be\label{anomdim}
\gamma=\frac{\lambda}{8\pi^2}\sum_i^{n_1}\frac{1}{(u_{1,i})^2+1/4}.
\ee

Given our restrictions
on the $R$-charges, we will only consider three cases.  These are ({\it i}\,)\ 
$n_2=n_3=n_1/2$ and so the representation is $[0,L-n_1,0]$, with 
$J_1=J$, $J_2=J_3=0$ and 
$n_1=L-J$.
({\it ii}\,)\ 
$n_2=n_1/2$, $n_3=0$, and so the representation is $[0,L-n_1-n_2,2n_2]$
with $J_1=J$, $J_2=J_3=J'$,   $n_2=J'$ and $n_1=L-J$.  
({\it iii}\,)\  $n_1=L/2+n_2/2$, $n_3=0$ and so the representation is
$[n_1-2n_2,0,n_1]$ with $J_1=J_2=J'$, $J_3=J$,  $J'=n_1-n_2$ and $J=n_2$.

In the first two cases we will be looking for the operator with the lowest anomalous dimension
for a given set of $R$-charges and bare dimension.  
For this reason the distribution
of the roots will be highly symmetric.  In the third case, we will not 
have the lowest anomalous dimension for the given representation, but we will
still have a symmetric distribution of roots. We shall see that the cases ({\it ii}\,)\ 
and ({\it iii}\,)\  are related by analytic continuation. In the course of the
analytic continuation a level crossing should occur where another branch
of semiclassical states becomes the global
minimum of the anomalous dimension in the $(J',J',J)$ sector.  These 
semiclassical states are the dual of the folded string when $J\to0$.

In the rest of this section we will consider cases ({\it i}\,) and ({\it ii}\,).  
Case ({\it iii}\,) will be discussed in a later section.

We proceed as in \cite{Beisert:2003xu}, 
where we assume  the number of roots is
of order  $L$.  We assume  the roots  are equally distributed about
$u=0$ and the distribution of $u_1$ roots
is of the same form as in \cite{Beisert:2003xu}, with the roots 
separated into two symmetric curves that intersect the real axis.
Taking logs of the equations in  \eqref{betheSO6}, rescaling
$u$ by a factor of $L$ and replacing sums  by integrals, we are left with the
following equations:
\begin{eqnarray}\label{inteqs}
\frac{1}{u}-2\pi m&=& \alpha\pint_{\CC_+}du'\frac{\s(u')}{u-u'}+\alpha\int_{\CC_+}du'\frac{\s(u')}{u+u'}
- \beta\int_{\CC_2}du'\frac{\rho_2(u')}{u-u'}- \beta'\int_{\CC_3}du'\frac{\rho_3(u')}{u-u'}\nonumber\\
0&=& 2\beta\pint_{\CC_2}du'\frac{\rho_2(u')}{u-u'}-\frac{\alpha}{2}\int_{\CC_+}du'\frac{\s(u')}{u-u'}-\frac{\alpha}{2}\int_{\CC_+}du'\frac{\s(u')}{u+u'}
\nonumber\\
0&=& 2\beta'\pint_{\CC_3}du'\frac{\rho_3(u')}{u-u'}-\frac{\alpha}{2}\int_{\CC_+}du'\frac{\s(u')}{u-u'}-\frac{\alpha}{2}\int_{\CC_+}du'\frac{\s(u')}{u+u'}
\end{eqnarray}
where $\alpha=n_1/L$, $\beta=n_2/L$ and $\beta'=n_3/L$.
$\CC_+$ is the right contour for the $u_1$ roots, $\CC_2$ is the contour
for the $u_2$ roots and $\CC_3$ is the contour for the $u_3$ roots.  The left
contour of the $u_1$ roots, $\CC_-$, is assumed to be the mirror image of
$\CC_+$.  The root densities are normalized to 
\be\label{rootdens}
\int_{\CC_+}\s(u')du'=\int_{\CC_2}\rho_2(u')du'=\int_{\CC_3}\rho_3(u')du'=1.
\ee

If we think of the Bethe roots as corresponding to the positions
of three different types of particles, then 
the solutions to the integral equations in \eqref{inteqs} 
give their equilibrium positions.  From these equations we see
that particles
of the same type repulse each other.  Also, the first type of
particles are attracted to the other two types and are
 also  attracted to a potential that has a minimum
at $u=\pm(2\pi m)^{-1}$.  The particles of the second and third type do
not see the
potential, nor do they interact directly with
each other.  Assuming  
the particles of the first type are arranged in two equal curves intersecting
the real axis, then the
particles of the second and third type must lie
 on the imaginary axis. 

We can thus solve for $\rho_2(u)$ and $\rho_3(u)$ in terms of $\s(u)$ 
and substitute the result back into the first equation in \eqref{inteqs}.
Performing Hilbert transforms, we find that
\be
\rho_2(iu)=-\frac{\alpha}{2\pi^2\beta}\sqrt{c^2-u^2}
\pint_{-c}^{c}\frac{du'}{u-u'} \frac{u'}{\sqrt{c^2-u^2}}\int_{\CC_+}du''\frac{\s(u'')}{(u')^2+(u'')^2}
\ee
and a similar equation for $\rho_3$.  Inverting the order of integration
and deforming the contour, we find
\be\label{rho2eq}
\rho_2(iu)=\frac{\alpha}{2\beta\pi}\int_{\CC_+} du''\frac{\s(u'')u''}{u^2+(u'')^2}\frac{\sqrt{c^2-u^2}}{\sqrt{c^2+(u'')^2}}
\ee
To determine $c$, we plug \eqref{rho2eq} into \eqref{rootdens}, where we find
\begin{eqnarray}
\int_{-c}^{+c}du'\rho_2(iu')&=&\frac{\alpha}{2\beta}\int_{\CC_+}du'\s(u')
-\frac{\alpha}{2\beta}\int_{\CC_+}du'\frac{\s(u')}{\sqrt{c^2+u'^2}}
\nonumber\\
&=&\frac{\alpha}{2\beta}-\frac{\alpha}{2\beta}\int_{\CC_+}du'\frac{\s(u')}{\sqrt{c^2+u'^2}}=1.
\end{eqnarray}
If $\beta=\alpha/2$, then
 $c=\infty$.  Assuming this ``half-filling'' condition,  one can
easily show that
\be
u\int_{-\infty}^{+\infty}du'\frac{\rho_2(iu')}{u^2+(u')^2}=\int_{C_+}du'
\frac{\s(u')}{u+u'}\ .
\ee
It is also clear that if the  $u_3$ roots are  half-filled, then
$\rho_3(u)$ satisfies the 
same equation.  Hence, if both sets of roots are half-filled
then this is case ({\it i}\,) and the first integral equation in
\eqref{inteqs} reduces to 
\be\label{case1ie}
\frac{2}{\alpha}\left(\frac{1}{u}-2\pi m\right)\ =\ 2\pint_{\CC_+}du'\frac{\s(u')}{u-u'}\ .
\ee
If the $u_2$ roots are half-filled and there are no $u_3$
roots then this is case ({\it ii}\,) and the first  integral equation in
\eqref{inteqs} reduces to 
\be\label{case2ie}
\frac{2}{\alpha}\left(\frac{1}{u}-2\pi m\right)\ =\ 2\pint_{\CC_+}du'\frac{\s(u')}{u-u'}\ +\ \int_{\CC_+}du'\frac{\s(u')}{u+u'}\ .
\ee

The eigenvalue density for an $O(n)$ matrix model satisfies the integral 
equation \cite{Kostov:fy,Gaudin:vx}
\be\label{Oneq}
U'(u)=2\pint_b^a du'\frac{\s(u')}{u-u'}-n\int_b^a du'\frac{\s(u')}{u+u'}.
\ee
If we compare \eqref{Oneq} to \eqref{case1ie} and \eqref{case2ie}, we see that 
these are both of this form with
\be\label{poteq}
U'(u)=\frac{2}{\alpha}\left(\frac{1}{u}-2\pi m\right).
\ee
and $n=0$ for case ({\it i}\,) and $n=-1$ for case ({\it ii}\,).

\sectiono{The gauge dual of the pulsating string}

We start with case ({\it i}\,) since this is simpler.  This has only
one $R$-charge $J$ and a bare dimension $L>J$.  We claim that this
is the SYM dual to the pulsating string.  

We first do a Hilbert 
transform on $\s(u)$ in \eqref{case1ie}, giving
\be\label{sigma1}
\s(u)\ =\ -\ \frac{\sqrt{(a-u)(u-b)}}{\pi\alpha u\sqrt{ab}},
\ee
where $a$ and $b$ are the end points of the cut.
These can be determined from \eqref{rootdens} which gives
\be
\frac{a+b}{\sqrt{ab}}=2(1-\alpha),
\ee
 and by explicitly plugging
\eqref{sigma1} into \eqref{case1ie}, which gives
\be
\pint_{b}^a du'\frac{\s(u')}{u-u'}=\frac{1}{\alpha}\left(\frac{1}{u}-
\frac{1}{\sqrt{ab}}\right).
\ee
Hence we have
\be\label{abeq}
\sqrt{ab}=\frac{1}{2\pi m}\qquad\qquad a+b=\frac{1}{\pi m}(1-\alpha).
\ee
For negative $\alpha$, we see that $a$ and $b$ are both real.  When
$\alpha$ is positive then $b=a^*$.

In order to find the anomalous dimension $\gamma$ it is convenient to
define the resolvent $W(u)$
\be\label{resolv}
W(u)\ =\ \int_b^a du'\frac{\s(u')}{u-u'}.
\ee
Using \eqref{sigma1} and \eqref{abeq}, we find that
\be\label{resolv2}
W(u)=\frac{1}{u}\left(1-\sqrt{(1-2\pi mu)^2+2\alpha(2\pi mu)}\right)-2\pi m.
\ee 
{}From \eqref{anomdim}, \eqref{resolv} and \eqref{resolv2} we see that $\gamma$ is given by
\be\label{adres}
\gamma\ =\ -\ \frac{\la}{8\pi^2L}\alpha W'(0)=
\frac{\lambda m^2}{4L}\alpha(2-\alpha),
\ee
precisely matching the result from the previous section.

We can say a bit more about our result.  First, there is a critical point
at $\alpha=1$.  This is the point where the representation is the $SO(6)$
singlet, so it is not surprising to find the critical behavior here.
At this critical point, we find that $a=-b=\frac{i}{2\pi m}$.  Hence at
this value, the contour $\CC_+$ is touching the imaginary axis and its
mirror $\CC_-$.

The simplicity of the root distribution also makes it easy to consider the
higher conserved charges.  
The generator of higher charges  is  \cite{faddeev}
\be\label{gencharges}
t(u)\ =\ \sum_n t_n u^n\ =\ i \log\left(\prod_k^{n_1}\frac{u-u_{1,k}+i/2}{u-u_{1,k}-i/2}\right),
\ee
Under the rescaling, this reduces to
\be\label{charges}
t(u)\ =\ -\ \frac{\alpha}{2}\left(W(u)-W(-u)\right),
\ee
and the charges $t_n$ are rescaled by a factor of $L^{-n}$.
We have already seen that $\gamma$ is related to the linear coefficient in
$t(u)$.  The next nontrivial charge is
\be
t_3=\frac{(2\pi m)^4}{8L^3}\alpha(2-\alpha)(5(1-\alpha)^2-1).
\ee
What this corresponds to on the string side is not immediately clear.

Even though half the charges are zero, we can still glean some information
from them.  For example, the total momentum on the string must be
zero because of level matching.  In SYM, this 
corresponds to the cyclicity property of the trace.  But we can still
determine the left and right moving contribution  to the momentum.
This is just
\be
-\frac{\alpha L}{2}W(0)=2\pi m (\alpha L/2).
\ee
In other words, the left moving momentum is $2\pi m$ multiplied by
half the number of impurities.  This is the result in the BMN limit,
so for this configuration, there is no correction
even with  a large number of impurities.

We can also go back and compute $\rho_2(iu)=\rho_3(iu)$, where we find
\be
\rho_2(iu)=\frac{2m}{\alpha}\left(1+\frac{\sqrt{(a-iu)(b-iu)}}{2iu}-\frac{\sqrt{(a+iu)(b+iu)}}{2iu}\right).
\ee
In the limit  $\alpha\to 1$, this reduces to
\begin{eqnarray}
\rho_2(iu)&=&2m\qquad\qquad\qquad\qquad\qquad -\frac{1}{2\pi m}\le u\le\frac{1}{2\pi m}
\nonumber\\
          &=&2m\left(1-\sqrt{1-(2\pi m u)^{-2}}\right) 
\qquad |u|>\frac{1}{2\pi m}.
\end{eqnarray}
This is the same distribution of $u_2$ and $u_3$ roots  found 
in \cite{Beisert:2003xu} for an $SO(6)$ singlet.  In this
case the $u_1$ roots were also on the imaginary axis with a density
twice that of the other two roots.
Comparing the $\alpha=1$ result for $t(u)$ in \eqref{charges} to 
the corresponding generator for the imaginary root solution, one
finds that they match.
This suggests that imaginary root Bethe state is equivalent
to the $\alpha=1$ Bethe state here.  

\sectiono{The SYM dual of the Frolov-Tseytlin string (I)}

Let us now consider case ({\it ii}\,) with equation \eqref{case2ie}.
This is the SYM dual to the Frolov-Tseytlin string \cite{Frolov:2003qc}
when $J'<J$.
In order to solve this, we use the results in
\cite{Eynard:1992cn}.  

The resolvent $W(u)$ is analytic everywhere except across the cut on
$\CC_+$.    On the cut, it follows from \eqref{case2ie}
that the resolvent satisfies the equation
\be\label{cuteq}
W(u+i0)+W(u-i0)-W(-u)=U'(u)
\ee
A solution to this equation is
\be\label{Wreq}
W_r(u)=\frac{1}{3}(2U'(u)+U'(-u))=\frac{2}{3\alpha\ u}-\frac{4\pi m}{\alpha}\ .
\ee
We will thus assume that
\be
W(u)=W_r(u)+w(u)
\ee
where $w(u)$ satisfies the homogeneous equation
\be
w(u+i0)+w(u-i0)=w(-u).
\ee
If we now consider the function $r(u)$,
\be\label{rdef}
r(u)\equiv w^2(u)-w(u)w(-u)+w^2(-u)\ ,
\ee
then it is simple to show that $r(u)$ is an even function
that is regular across the cut.
Given the form of $W_r(u)$ in \eqref{Wreq}, $r(u)$ approaches a constant
at infinity and has a double pole at $u=0$.  The form is easily determined
by recalling that $W(u)$ is regular at $u=0$, and falls off as $1/u$ for
large $u$.  Thus, $w(u)$ must be chosen to cancel off the constant piece
of $W_r(u)$ for large $u$ and the pole at $u=0$, as well as matching onto
the correct asymptotic behavior for $W(u)$.   This gives us 
\begin{eqnarray}\label{wasymp}
w(u)&\approx& \frac{4\pi m}{\alpha}+\left(1-\frac{2}{3\alpha}\right)\frac{1}{u}\qquad\qquad
u\to\infty
\nonumber\\
w(u)&\approx&-\frac{2}{3\alpha u}\qquad\qquad\qquad\qquad\qquad u\to 0\ ,
\end{eqnarray}
  and hence, we find 
\be\label{req}
r(u)=\frac{(4\pi m)^2}{\alpha^2}+\frac{4}{3\alpha^2u^2}\ .
\ee

If we now multiply $r(u)$ by $(w(u)+w(-u))$,  we find the equation
\be\label{cubeq}
w^3(u)-r(u)w(u)=-w^3(-u)+r(-u)w(-u)\equiv s(u)\ .
\ee
The function $s(u)$ is clearly an odd function, and since $w(u)$ is
regular for ${\rm Re}(u)<0$, and $w(-u)$ is regular for ${\rm Re}(u)>0$,
$s(u)$ is analytic everywhere except at $u=0$.  
Using \eqref{wasymp} and
\eqref{req}, we find that $s(u)$ is determined and is given by
\be
s(u)=\frac{16}{27\alpha^3u^3}+2\frac{(4\pi m)}{\alpha^2}\left(1-\frac{2}{3\alpha}\right)\frac{1}{u}\ .
\ee

Hence, finding the resolvent has been reduced to solving a cubic equation.
We do not actually need to do this, since to find the charges we
only need to expand about $u=0$.   Given the generating function $t(u)$ 
in \eqref{charges}, it is convenient to define the difference
\be
\bw(u)=w(u)-w(-u).
\ee
Then it is simple to show using \eqref{rdef} and \eqref{cubeq} that
$\bw(u)$ satisfies the cubic equation
\be\label{cubeq2}
\bw^3(u)-r(u)\bw(u)+s(u)=0.
\ee
Solving \eqref{cubeq2} as a series
expansion, we find that the generator of the charges $t(u)$ is
\begin{eqnarray}
t(u)&=&-\frac{\alpha}{2}\left(\bw(u)+W_r(u)-W_r(-u)\right)
\nonumber\\
&=&
\alpha(2\pi m)^2 u +\alpha(1-2\alpha)(2\pi m)^4u^3+\alpha(1-6\alpha+7\alpha^2)(2\pi m)^6u^5+...
\end{eqnarray}
In particular, we note that 
\be
W'(0)=\frac{1}{2}\bw'(0)=-(2\pi m)^2,
\ee
and so
\be
\gamma=\frac{\lambda m^2}{2L}\alpha=\frac{\lambda m^2J'}{L^2}.
\ee
This is the same result obtained by Frolov and Tseytlin from the 
semiclassical string \cite{Frolov:2003qc,Frolov:2003tu}!

It is interesting to find the end points of the cut.  At these points,
$W(u)$ and hence $w(u)$ has a singularity.  We can find these
points by looking for the zeroes of the discriminant of \eqref{cubeq2},
which is 
\begin{eqnarray}
\Delta&=&4r^3(u)-27s^2(u)
\nonumber\\
&=&
64\frac{(2\pi m)^2}{\alpha^6u^4}\left(4(1-\alpha)-(27\alpha^2-36\alpha+8)(2\pi m)^2u^2+4(2\pi m)^4 u^4\right).
\end{eqnarray}
This has zeroes at
\be\label{ueq}
u=\pm\frac{\sqrt{16-72\alpha+54\alpha^2\pm2 i\sqrt{\alpha(8-9\alpha)^3}}}{4(2\pi m)}\ ,
\ee
with the four solutions corresponding to the endpoints of $\CC_+$ and $\CC_-$.
Note that the system  has an apparent critical point
 at $\alpha=8/9$, where the
end points of the cut hit the imaginary axis at
\be
u=\pm \frac{i}{2\pi m\sqrt{3}}\ ,
\ee
and the contours $\CC_+$ and $\CC_-$ touch each other.
For this value of $\alpha$ we have $J'=4J$.  For $\alpha>8/9$ the end points
move along the imaginary axis, with two of them reaching $u=0$ and
the other two reaching $u=\pm \frac{i}{4\pi m}$ when $\alpha=1$ and $J=0$.

Strictly speaking, the Bethe states with $\alpha>2/3$, which corresponds
to $J'>J$,
do not exist.  This is because one of the Dynkin indices
is negative in this region, and since the Bethe states are highest weights,
the state must have zero norm.  However, the anomalous dimension
as well as all the higher charges are analytic across this point, so
we believe that there will be another set of Bethe states with nonzero
norm with these exact charges.  This is sufficient since
for an integrable system, the
state is completely determined by the conserved charges.
In the next section we will find Bethe states which are
valid for at least for some of this region.

The conserved charges are also analytic across $\alpha=8/9$.  For
this reason we believe  the critical point is an artifact of
the Bethe ansatz.  In the next section we will 
 show that these charges match with charges
where the Bethe states exist,  further demonstrating that the
critical point is a fake.   In anticipation of this, 
consider \eqref{cubeq2} when $\alpha=1$.
In this case the solution for $\bw(u)$  simplifies dramatically to 
\begin{eqnarray}
\bw(u)&=&-\frac{1}{3u}-\frac{\sqrt{1+(4\pi m)^2u^2}}{u}
\nonumber\\\label{tucont}
t(u)&=&\frac{-1+\sqrt{1+(4\pi m)^2u^2}}{2u}\ .
\end{eqnarray}


\sectiono{The gauge dual of the Frolov-Tseytlin string (II)}

We now turn to  case ({\it iii}\,)  where $J'>J$. When $J=0$ we only need one
type of Bethe root to generate the Bethe state which means that this is 
a state for the Heisenberg
XXX spin chain. The Bethe state dual to the Frolov-Tseylin $J=0$
solution 
was constructed in \cite{Beisert:2003xu} which we briefly review.

The starting point is the solution of the Bethe equations 
\be
u_{1,1}=0,\ \ \ u_{1,2}=i/2,\ \ \ u_{1,3}=-i/2\ .
\ee
 The equation for $u_{1,1}$ is identically satisfied provided
$L$ is even. The equations for the other two  roots acquire the form
$0=0$ or $\infty=\infty$. Appropriate infinitesimal shifts from $\pm i/2$
balance the singularities and  renders the energy finite.
 In fact, a zero or a pole appears on
the right hand side of the Bethe equations each time a pair of
roots is separated by $\pm i$, which allows us to put extra roots on the imaginary
axis. The left hand side of the Bethe equations will then be exponentially large or
exponentially small in $L$. To compensate, the roots should be put exponentially
close to $i n/2$ with integer $n$'s.  This  produces small denominators on the
right hand side of the Bethe equations. 
If the number of the roots is macroscopic, the half-integer pattern breaks down at some critical 
$n\sim L$ \cite{Beisert:2003xu} 
and the rest of the roots split along the imaginary axis by distances larger 
than $1/2$. If we parameterize the roots by $u_{1,k}=iq_{1,k}L$, and introduce 
the density
\begin{equation}
\sigma(q)=\frac{2}{L}\sum_k\delta(q-q_{1,k}),
\end{equation} 
it will consist of two parts, the condensate with $\sigma(q)=4$
and two tails with  $\sigma(q)<4$. We write this as
\begin{equation}
\sigma(q)=\left\{
\begin{array}{ll}
4,&~~~-s<q<s,\\
\tilde{\sigma}(q),&~~~s<q<t,\\
\tilde{\sigma}(-q),&~~~-t<q<-s,\\
0,&~~~|q|>t,
\end{array}
\right.
\end{equation}
The Bethe equations unambiguously determine 
$\tilde{\sigma}(q)$ and the number of roots in the condensate, $4sL$. The parameter $t$ 
is then fixed by the normalization condition
\begin{equation}\label{normsigma}
\int dq\,\sigma(q)=2\alpha\ .
\end{equation}
We have chosen to normalize the density differently than in the previous
sections for later convenience, but $\alpha L$ is still
the number of $u_1$ roots. Also the relationship between the number of the
roots and the R-charges is different here.
Explicit formulae for thr density in terms of elliptic integrals can be found in 
\cite{Beisert:2003xu}.
The results simplify considerably at $\alpha=1/2$ where
 $t\rightarrow\infty$
and the density reduces to an algebraic function. 

We can always add  $u_{2,1}=0$
to an arbitrary configuration of $u_1$ roots. 
This gives a solution of 
the Bethe equations with $L$
replaced by $L+1$ \cite{Minahan:2002ve}. By numerically solving the 
Bethe equations for several configurations with  
a handful of $u_2$ roots, we observed that the $u_2$ roots cluster 
near zero on the imaginary axis. 
We therefore expect that in the thermodynamic limit the density of $u_2$ roots will differ from zero 
on an interval from $-v$ to $+v$ with $v<s$. 
As above, the density is defined by
\begin{equation}
\rho(q)=\frac{2}{L}\sum_k\delta(q-q_{2,k}),
\end{equation}
where $u_{2,k}=iq_{2,k}L$, and is normalized to
\begin{equation}\label{normrho}
\int dq\,\rho (q)=2\beta,
\end{equation}
with $\beta L$ equal to the number of $u_2$ roots.

Let us for the moment assume that the $R$-charges are arbitrary
but the SYM dual is still holomorphic.  In this case the number of
roots of the Bethe state are related
to the $R$-charges
by $J_1=(1-\alpha)L$, $J_2=(\alpha-\beta)L$ and
$J_3=\beta L$.  We will call the condition $J_1=J_2$ ``half-filling'', and
at half-filling we have that $2\alpha-\beta=1$.

Taking the $L\rightarrow\infty$ limit of the Bethe equations, 
the roots outside of the condensate satisfy the equations
\begin{eqnarray}
\frac{1}{q_{1,k}}&=&\frac{2}{L}\sum_{l\neq k}\frac{1}{q_{1,k}-q_{1,l}}-\frac{1}{L}
\sum_l \frac{1}{q_{1,k}-q_{2,l}}\ ,
\nonumber  \\*
0&=&\frac{2}{L}\sum_{l\neq k}\frac{1}{q_{2,k}-q_{2,l}}-\frac{1}{L}
\sum_l \frac{1}{q_{2,k}-q_{1,l}}\ ,
\end{eqnarray}
 or in terms of the densities,
\begin{eqnarray}
\label{eq1}
2q \pint_s^t dq'\,\tilde{\sigma }(q')\,\frac{1}{q^2-q'{}^2}
&=&
\frac{1}{q}+\frac{1}{2}\int_{-v}^{v}dq'\,\rho (q')\, \frac{1}{q-q'}
-4\ln\frac{q-s}{q+s},~~~s<q<t\ ,
\\* \label{eq2}
\pint_{-v}^{v}dq'\,\rho (q')\, \frac{1}{q-q'}
&=&
q \int_s^t dq'\,\tilde{\sigma }(q')\,\frac{1}{q^2-q'{}^2}
+2\ln\frac{s+q}{s-q}\ ,~~~~-v<q<v.
\end{eqnarray}
As in section 3, if we think of $q_{i,k}$ as coordinates of particles on a 
line subject to pairwise logarithmic
interactions, these equations 
describe their equilibrium distribution. The logarithmic terms in the equations correspond to 
the interaction with
the condensate in $\sigma (q)$. Roots of the same type repulse  each other and  roots
of different types attract. In particular, $q_{1}$ roots create an effective potential 
for $q_2$ roots which has a minimum at zero and which confines the $q_2$ roots 
around  the origin. 
This justifies our assumption about the functional form of the density
of the second type of roots $\rho (q)$. The expression for the 
anomalous dimension depends
only on the density of $q_1$ roots and
so has the same form as for $\rho=0$ \cite{Beisert:2003xu}, 
namely\footnote{One has to carefully
take into account the contribution of the condensate to derive this formula.}
\begin{equation}
\gamma =\frac{\lambda }{8\pi ^2L}\left(\frac{4}{s}-\int_s^tdq\,\frac{\tilde{\sigma }(q)}{q^2}\right).
\end{equation}

There is a strong resemblance between the thermodynamic limit of 
the Bethe equations
and saddle-point equations in large-$N$ matrix models. In particular, the equation (\ref{eq2}) is
the same as the saddle-point equation in the Hermitian one-matrix model
\cite{Brezin:1977sv},
while
(\ref{eq1}) arises in large-$N$ two-dimensional QCD on a sphere \cite{Douglas:1993ii
}. Our strategy 
 will be to find $\tilde{\sigma }$ from the first equation, treating $\rho $ as an
external field and then to solve for $\rho $.
The solution to (\ref{eq1}) with a infinitesimal $\rho $ was found in 
\cite{Beisert:2003xu} by adapting the  techniques of \cite{Douglas:1993ii}. 
Since the equation is linear it is not hard 
to write down 
the general solution,
\begin{eqnarray}
\tilde{\sigma }(q)&=&\frac{1}{\pi }\sqrt{(q^2-s^2)(t^2-q^2)}
\left[-\frac{1}{qst}+4\int_{-s}^s \frac{dx}{(q-x)\sqrt{(s^2-x^2)(t^2-x^2)}}
\right. \nonumber  \\*
&&\left.
-\frac{1}{2}\int_{-v}^v \frac{dx\,\rho (x)}{(q-x)\sqrt{(s^2-x^2)(t^2-x^2)}}\right].
\end{eqnarray}
By plugging this expression into eq.~(\ref{eq1}), we get an additional constraint
on $s$ and $t$,
\begin{equation}\label{cons}
4\int_{-s}^s \frac{dx}{\sqrt{(s^2-x^2)(t^2-x^2)}}
=\frac{1}{st}+\frac{1}{2}\int_{-v}^v \frac{dx\,\rho (x)}{\sqrt{(s^2-x^2)(t^2-x^2)}}\ .
\end{equation}
Another constraint is derived from the normalization conditions 
\eqref{normsigma} and \eqref{normrho},
\begin{equation}\label{normcond}
4\int_{-s}^s \frac{dx\, x^2}{\sqrt{(s^2-x^2)(t^2-x^2)}}
-\frac{1}{2}\int_{-v}^v \frac{dx\,\rho (x)x^2}{\sqrt{(s^2-x^2)(t^2-x^2)}}
=1-2\alpha +\beta .
\end{equation}
Since $\rho(x)<4$, the left hand side of \eqref{normcond}
is manifestly positive.
Half-filling $1-2\alpha+\beta=0$ therefore corresponds to a critical
point at which $t$ goes to infinity. The density then simplifies to
\begin{equation}
\tilde{\sigma }(q)=\left(4-\frac{1}{\pi s}\sqrt{1-\frac{s^2}{q^2}}-\frac{1}{2\pi }
\int_{-v}^v dx\,\frac{\rho (x)}{q-x}\sqrt{\frac{q^2-s^2}{s^2-x^2}}\right)
\end{equation}
and the consistency condition \eqref{cons} becomes
\begin{equation}\label{con}
4\pi =\frac{1}{s}+\frac{1}{2}\int_{-v}^v \frac{dx\,\rho (x)}{\sqrt{s^2-x^2}}.
\end{equation}
This condition ensures that the density decreases at infinity.
We can also compute the anomalous dimension which is given by
\begin{equation}
\gamma =\frac{\lambda }{32\pi ^2L}\left[\frac{1}{s^2}
+\int_{-v}^vdq\,\rho (q)\,\frac{1}{q^2}\left(\frac{s}{\sqrt{s^2-q^2}}-1\right)\right].
\end{equation}

The next step is to substitute $\tilde{\sigma }$ into (\ref{eq2}) and to solve for
$\rho (q)$. We then find
\begin{equation}
\pint_{-v}^v\frac{dx\,\rho (x)}{q-x}\left(3+\sqrt{\frac{s^2-q^2}{s^2-x^2}}\right)
=\frac{2}{q}\left(1-\sqrt{1-\frac{q^2}{s^2}}\right).
\end{equation}
This equation has a simpler analytic structure than it may seem 
because the square roots
 can be eliminated by a simple change of variables:
\begin{equation}
q=\frac{2s\eta }{1+\eta ^2},~~~~~x=\frac{2s\xi }{1+\xi ^2},~~~~~~dx\rho (x)=d\xi \rho (\xi ).
\end{equation}
In the new variables the integral equation is
\begin{equation}\label{inteq}
\pint d\xi \,\rho (\xi )\,\frac{1+\xi ^2}{1-\xi ^2}\left(2\,\frac{1+\eta \xi }{\eta -\xi }
+\frac{\eta +\xi }{1-\eta \xi }\right)=2\eta ,
\end{equation}
where it  contains only rational coefficients.

The consistency condition (\ref{con}) then reads
\begin{equation}\label{cc}
\int d\xi \,\rho (\xi )\,\frac{1+\xi ^2}{1-\xi ^2}=8\pi s-2,
\end{equation}
and the anomalous dimension is
\begin{equation}
\gamma =\frac{\lambda }{32\pi ^2Ls^2}\left[
1+\frac{1}{2}\int d\xi \,\rho (\xi )\,\frac{(1+\xi ^2)^2}{1-\xi ^2}\right].
\end{equation}

After
the further change of variables, $\xi =(1-p)/(1+p)$, \eqref{inteq}
reduces to
the $O(n)$ matrix model form in \eqref{Oneq}, but now with $n=1$. 
However, we found it more practical to
do the calculation in the original variables while  repeating
 the  steps in \cite{Eynard:1992cn}. 
To this end, let us define the resolvent
\begin{equation}
F(z)=\int d\xi \,\rho (\xi )\,\frac{1+\xi ^2}{1-\xi ^2}\,\frac{1+z\xi }{z-\xi }\,.
\end{equation}
$F(z)$ is an odd function whose only singularities are a pole at infinity
given by
\begin{equation}\label{boundr}
F(z)=\frac{p}{z}+\ldots,~~~~~(z\rightarrow\infty),
\end{equation}
and a branch cut from
$-\nu $ to $\nu $, where $\nu $ is related to $v$ by
\begin{equation}
v=\frac{2s\nu }{1+\nu ^2}.
\end{equation}
 The branch points  lie inside the unit circle because 
$v$ is  smaller than $s$.
When the density of $u_2$ roots approaches the end-point of the condensate, 
$\nu $ goes to one. The residue at infinity is given by
$$
p=\int d\xi \,\rho (\xi )\,\frac{(1+\xi ^2)^2}{1-\xi ^2},
$$
and hence
\begin{equation}
\gamma =\frac{\lambda }{32\pi ^2Ls^2}\left(1+\frac{p}{2}\right).
\end{equation}
The consistency condition (\ref{cc}) and the normalization \eqref{normrho} 
can be easily expressed in terms of $F(z)$
\begin{eqnarray}
F(i)&=&-i(8\pi s-2),\label{co1}
\\*
F'(i)&=&2\beta . \label{co2}
\end{eqnarray}

The resolvent satisfies a functional equation which can be derived by
multiplying both sides of (\ref{inteq}) by 
$$
\rho (\eta )\, \frac{1+\eta ^2}{1-\eta ^2}\left(\frac{1}{z-\eta }-\frac{1}{z-1/\eta }\right)
$$
and integrating over $\eta $. A long but straightforward calculation yields
\begin{equation}
F^2(z)+F^2(1/z)+F(z)F(1/z)-2zF(z)-(2/z)F(1/z)+64\pi ^2s^2-4=0.
\end{equation}
In order to get rid of the linear terms,
we expand $F(z)$ as
\begin{equation}\label{defw}
F(z)=\frac{4z}{3}-\frac{2}{3z}+w(z).
\end{equation}
$w(z)$ satisfies the purely quadratic equation
\begin{equation}\label{222}
w^2(z)+w^2(1/z)+w(z)w(1/z)=R(z),
\end{equation}
where
\begin{equation}
R(z)=R(1/z)=\frac{4}{3}(z+1/z)^2-64\pi ^2s^2.
\end{equation}
Multiplying  (\ref{222}) by $w(z)-w(1/z)$, we again  find a cubic equation
\begin{equation}\label{quibic}
w^3(z)-R(z)w(z)=w^3(1/z)-R(1/z)w(1/z)\equiv S(z).
\end{equation}
The manifestly odd function $S(z)$ is symmetric under $z\rightarrow 1/z$, has poles at zero
and at infinity and potentially has branch points at $\pm \nu $, $\pm 1/\nu $. But
the left hand side of (\ref{quibic}) is analytic at $\pm 1/\nu $ and the middle
is analytic at $\pm \nu $. Consequently, $S(z)$ has no branch points, its only singularities
are poles at zero and at infinity, therefore it is an odd polynomial in $z+1/z$ of at most
third degree. Taking into account the definition of $w(z)$, eq.~(\ref{defw}),
and the boundary condition (\ref{boundr}), we get
\begin{equation}
S(z)=-\frac{16}{27}(z+1/z)^3+\frac{4}{3}(6+3p-64\pi ^2s^2)(z+1/z).
\end{equation}
Solving the cubic equation (\ref{quibic}), we can find $w(z)$ and therefore $F(z)$.
However, we do not need the explicit form of the resolvent to compute the anomalous 
dimension. We only need to know the constant $p$ and that is determined by the
constraints (\ref{co1}), (\ref{co2}). Putting $z=i$ in (\ref{quibic}) and taking into account
(\ref{co1}) gives an identity and does not lead to any relation between the 
parameters. But if we first differentiate in $z$ and then put $z=i$, we find 
the non-trivial
equation for $p$ in terms of $s$ and $\beta $,
\begin{equation}
p=32\pi ^2s^2(1-\beta )-2.
\end{equation}
Therefore, the anomalous dimension again has the amazingly simple form
\begin{equation}
\gamma =\frac{\lambda (1-\beta )}{2L}=\frac{\lambda J'}{L^2}\,,
\end{equation}
which agrees with the string-theory prediction of Frolov and 
Tseytlin \cite{Frolov:2003qc},
and is consistent with the solution in the previous section.

Let us now see how the number of roots in the condensate depends on the total
number of $u_2$ roots, {\it i.e.}  the relation of  $s$ to  $\beta $. 
There are no
other constraints or boundary conditions imposed on the resolvent $F(z)$
than those that we have already used to find the anomalous dimension, so the
dependence of $s$ on $\beta $ must be determined by the analytic structure of $F(z)$.
Like the previous section, we examine the discriminant of the cubic equation (\ref{quibic}),
\begin{equation}
\Delta (z)=4R^3(z)-27\,S^2(z).
\end{equation}
The solution of the cubic equation and therefore the resolvent $F(z)$ depends on $z$
through $\sqrt{\Delta (z)}$. Single zeros of the discriminant (but not double zeros!)
are branch points of the resolvent.  Using the explicit expressions for
$R(z)$ and $S(z)$ we find
\begin{equation}
\Delta =-4096\pi ^2s^2\beta \left[(z+1/z)^2-2\pi ^2s^2\chi _+/\beta \right]
\left[(z+1/z)^2-2\pi ^2s^2\chi _-/\beta \right],
\end{equation}
where
\begin{equation}
\chi _{\pm}=1+18\beta -27\beta ^2\pm(1-9\beta )\sqrt{(1-9\beta )(1-\beta )}\,.
\end{equation}
The discriminant $\Delta $, as a function of $z+1/z$, has four zeros. Consequently, 
the resolvent will have four branch points instead of two, 
unless $2\pi ^2s^2\chi _-/\beta =4$
in which case the discriminant has a double zero at $z=\pm 1$.   Insisting on
only two branch points we find 
\begin{equation}\label{seq}
s=\frac{\sqrt{2+36\beta -54\beta ^2 +2\sqrt{(1-\beta )(1-9\beta )^3}}}{8\pi }\,.
\end{equation}
Examining \eqref{seq} we see that $s$ grows with $\beta$
because $u_1$ roots in the tail of the distribution
 are attracted toward the $u_2$ roots at the origin. The parameter $s$ changes from
$1/(4\pi )$ at $\beta =0$ to $1/(2\pi\sqrt{3} )$ at $\beta =1/9$ and becomes complex for larger
$\beta $. But $s$ is real by definition, hence $\beta =1/9$ is a critical 
point for the Bethe equations. In terms of $R$-charges, $\beta =1/9$ corresponds to
$J'=4J$, the point we found in section 5.

Let us now
look at the branch points of the resolvent:
\begin{equation}\label{veq}
v=\frac{\sqrt{2+36\beta -54\beta ^2 -2\sqrt{(1-\beta )(1-9\beta )^3}}}{8\pi }\,.
\end{equation}
As $\beta \rightarrow 1/9$, $v\rightarrow s$ and the density 
$\rho $ of $u_2$ roots collides with the tail
of the distribution of  $u_1$ roots, $\tilde{\sigma }$.
This type of critical behavior corresponds to the Ising phase transition
in the $O(1)$ matrix model \cite{Eynard:1992cn}.   But in the case of the string,
we believe that it is only a signal that this particular configuration
of Bethe roots can no longer describe the string state and that 
the physical system happily continues through this point.  Indeed,
if we compare \eqref{seq} and \eqref{veq} with the end points of the
cuts in \eqref{ueq}, and recall that $\alpha$ in \eqref{ueq} 
is $2J'/L$ and is related to $\beta $ by $\alpha =1-\beta $, 
we see that the end points of $\CC_+$ match onto $s$ and $-v$
and the end points of $\CC_-$ match onto $-s$ and $v$.  This strongly
suggests that the analytic continuation of the $J'<J$ sector is
the sector described in this section, or at least if $J'>4J$.

As a further check on the spurious nature of the critical point let us consider
the higher conserved charges when $J=0$, which is in the $J'>4J$ region.
The contribution of the condensate to $t(u)$ in \eqref{gencharges}, after
rescaling, is
\be
t_c(u)=i\prod_{n=-2sL}^{2sL}\log\left(\frac{uL-ni/2+i/2}{uL-ni/2-i/2}\right)
=2i\log\frac{u+is}{u-is}\ .
\ee
Using the normalization condition in \eqref{normsigma}, one finds 
\be\label{tuJ0}
t(u)=2i\log\frac{u+is}{u-is}-\frac{1}{2}
\left(\int_{-\infty}^{-s}dq\frac{\ts(q)}{u-iq}
+\int_s^\infty dq\frac{\ts(q)}{u-iq}\right)\ .
\ee
In \cite{Beisert:2003xu} it was shown that
\be\label{tseq}
\ts(q)=4(1-\sqrt{1+s^2/q^2})
\ee
when $J=0$.  Inserting \eqref{tseq} into \eqref{tuJ0} and deforming the 
contour,  leads to
\be
t(u)=2\pi  \left(-\frac{s}{u}+\sqrt{1+s^2/u^2}\right)\ .
\ee
Since $s=1/4\pi$ when $J=0$, we see that this is the $m=1$ result in
\eqref{tucont}.  Hence, all conserved charges for the $J'<J$ Bethe states
 continue through onto
the $J=0$ Bethe state.

\sectiono{Discussion}

Combining the results of this paper with \cite{Beisert:2003xu},
we can summarize the SYM duals of different semiclassical string motion
as follows:  The folded string is dual to an $O(-2)$ model, the
circular string is dual to an $O(-1)$ or $O(+1)$ model and the
pulsating string is dual to an $O(0)$ model.  A natural question
to ask is whether other $O(n)$ models are related to yet other types
of semiclassical string motion.   One might also wonder if there
is a deeper connection between the strings in $AdS_5\times S^5$ and the 
matrix models.  For
example, the $O(n)$ models are known to be critical only within the
range $-2<n<2$, hence $-2$ is a limiting value.  The same is true of
the folded string, in the sense that this is a limit of an ellipsoidal
string that smoothly interpolates into a circular string.

Another interesting question concerns the stability of the semiclassical
strings.  In \cite{Frolov:2003qc} it was shown that the single
wound circular string develops an unstable mode at $J'\ge 3J/2$.  Presumably
this can be checked on the dual side, where one can look for the spinless
modes  by shifting roots around.  This was accomplished for the $J=0$
case in \cite{Beisert:2003xu} and we expect the same techniques to
work when $J\ne0$.

It would also be interesting to extend these results to $\alpha<0$.  This
was done in \cite{Beisert:2003ea} to the results of
\cite{Beisert:2003xu}.  The authors showed that the resulting integral
equation was the thermodynamic limit of a particular sector
of the $SU(2,2|4)$ Bethe equations
in \cite{Beisert:2003yb}.  They then went on to show that anomalous
dimension matched the predicted anomalous dimension for a folded string
spinning in $AdS_5$ and with angular momentum in $S_5$ 
\cite{Frolov:2002av}.  We expect
a similar phenomenon to occur here.

Another worthwhile goal is to better understand the higher
charges from the point of view of the semiclassical string.
It is not immediately clear what these conserved charges would mean.
The semiclassical string contains information about all orders
in $\lambda$, but the integrability of the dilatation
operator has been established only at one-loop, although there are
hints that integrability can be taken further in SYM 
\cite{Beisert:2003tq,Beisert:2003jb}, as well as cautions \cite{Callan:2003xr}.
Nevertheless, it has been recently pointed out that the $AdS_5\times S_5$
sigma model has an infinite tower of non-local conserved charges 
\cite{Bena:2003wd,Mandal:2002fs}.  This was shown to also hold in the Berkovits 
description \cite{Vallilo:2003nx} and was further analyzed in the
plane-wave limit \cite{Alday:2003zb}.  In \cite{Dolan:2003uh}, some
progress was made toward relating the integrability of the one-loop
dilatation operator and the non-local symmetries in the sigma model.
Hopefully, the elegant yet simple equation in \eqref{charges}
relating the generator of conserved
charges to the resolvent
of the Bethe roots
will provide further clues.

\bigskip
\vfill\eject

{\it Note added:} After completion of this work we learned about 
work of Arutyunov and Staudacher \cite{Arutyunov:2003rg} 
 where they show that higher charges
coincide on both sides of the AdS/CFT correspondence for the solutions
in \cite{Frolov:2003qc,Frolov:2003xy,Beisert:2003xu}.

\bigskip\bigskip
\noindent {\bf Acknowledgments}:
We would like to thank G.~Arutyunov.
 N.~Beisert, V.~Kazakov and M.~Staudacher for discussions
and V.~Kazakov for bringing ref.~\cite{Eynard:1992cn} to our attention.
This research was
supported in part by the Swedish Research Council. The research
of J.A.M. was also supported in part by DOE contract \#DE-FC02-94ER40818.
The research of K.Z. was also supported in part by
RFBR grant 02-02-17260 and in part by RFBR grant
00-15-96557 for the promotion of scientific schools.

\end{document}